\documentclass[11pt,letter]{article}

\usepackage{tabularx}
\usepackage{amsmath} 
\usepackage{hhline} 
\usepackage{url}

\usepackage{hyperref}
\usepackage[numbers]{natbib}
\usepackage{graphicx}
\usepackage{times}

\newcommand{\hmmerO}{\textsf{HO}}
\newcommand{\hmmerM}{\textsf{HB}}
\newcommand{\hmmerG}{\textsf{HG}}
\newcommand{\psiblastO}{\textsf{PO}}
\newcommand{\psiblastC}{\textsf{PC}}
\newcommand{\psiblastS}{\textsf{PS}}

\newcommand{\hmmerOF}{\textsf{HOF}}
\newcommand{\hmmerMF}{\textsf{HBF}}
\newcommand{\hmmerGF}{\textsf{HGF}}
\newcommand{\psiblastOF}{\textsf{POF}}
\newcommand{\psiblastCF}{\textsf{PCF}}

\newcommand{\hmmerOU}{\textsf{HOU}}
\newcommand{\hmmerMU}{\textsf{HBU}}
\newcommand{\hmmerGU}{\textsf{HGU}}
\newcommand{\psiblastOU}{\textsf{POU}}
\newcommand{\psiblastCU}{\textsf{PCU}}
\newcommand{\psiblastSU}{\textsf{PSU}}

\newcommand{\ROC}[1]{\text{ROC}_{#1}}
\newcommand{\RRSD}[1]{\text{RRSD}_{#1}}

\def\be{\begin{equation}}
\def\ee{\end{equation}}
\def\bea{\begin{eqnarray}}
\def\eea{\end{eqnarray}}
\def\To{\!\to\!}

\def\ba{\mathbf a}
\def\bb{\mathbf b}

\begin{document}
\begin{titlepage}

\begin{center}
{\Large\bf The effectiveness of position- and composition-specific gap costs for protein similarity searches}
\end{center}
\vspace{.35cm}

\begin{center}
{\large Aleksandar Stojmirovi\'c,\,  E. Michael Gertz,\, Stephen F. Altschul,\ and Yi-Kuo Yu\footnote{to whom correspondence should be addressed}}
\vspace{0.25cm}
\small

\par \vskip .2in \noindent
National Center for Biotechnology Information\\
National Library of Medicine\\
National Institutes of Health\\
Bethesda, MD 20894\\
United States
\end{center}

\normalsize
\vspace{0.25cm}

\begin{abstract}

\subsubsection*{Motivation:}
The flexibility in gap cost enjoyed by Hidden Markov Models (HMMs) is expected to afford them better retrieval accuracy than position-specific scoring matrices (PSSMs). We attempt to quantify the effect of more general gap parameters by separately examining the influence of position- and composition-specific gap scores, as well as by comparing the retrieval accuracy of the PSSMs constructed using an iterative procedure to that of the HMMs provided by Pfam and SUPERFAMILY, curated ensembles of multiple alignments. 

\subsubsection*{Results:}
We found that position-specific gap penalties have an advantage over uniform gap costs. We did not explore optimizing distinct uniform gap costs for each query. For Pfam, PSSMs iteratively constructed from seeds based on HMM consensus sequences perform equivalently to HMMs that were adjusted to have constant gap transition probabilities, albeit with much greater variance. We observed no effect of composition-specific gap costs on retrieval performance. 

\subsubsection*{Availability:}
The scripts for performing evaluations are available upon request from the authors.

\subsubsection*{Contact:} \href{yyu@ncbi.nlm.nih.gov}{yyu@ncbi.nlm.nih.gov}
\end{abstract}
\end{titlepage}

\section{Introduction}

Information retrieval from molecular databases by sequence alignment is an essential component of modern biology. The effectiveness of retrieval strategies depends  crucially on how alignments are scored. A pairwise alignment score typically combines scores for the substitutions, insertions, and deletions that transform one sequence into another. Scores for substitutions are derived from a substitution matrix, while scores for insertions and deletions are known as gap costs. The importance of gap costs has prompted numerous studies proposing various reasonable gap penalty schemes \citep{PA92,BCG93,RP02,CB04,WG04,QE06}.

Search accuracy may be improved substantially by using position-specific scoring matrices (PSSM) \citep{GME87}. In addition, it is possible to introduce position- and composition-specific gap costs, which so far have been implemented primarily by Hidden Markov Models (HMMs) \citep{KBMSH94,DEKM98}. In this paper we attempt to quantify the effect of different gap scores on retrieval performance using PSI-BLAST \citep{AMSZZML97} and HMMER \citep{Eddy98,Eddy03}, canonical examples of software tools employing PSSMs and HMMs, respectively.

As its name suggests, a PSSM assigns scores to amino acids in a database sequence based on the position in which they occur in the alignment. PSI-BLAST computes and scores alignments using a heuristic approximation to the Smith-Waterman algorithm \citep{SW81} with affine gap costs \citep{Go82} providing uniform penalties for opening and extending a gap. PSSMs used by PSI-BLAST may be generated through an iterative search procedure, or obtained from other sources, such as databases of curated multiple sequence alignments (MSAs).

\begin{figure}[h!]
\begin{center}
\scalebox{0.65}{\includegraphics{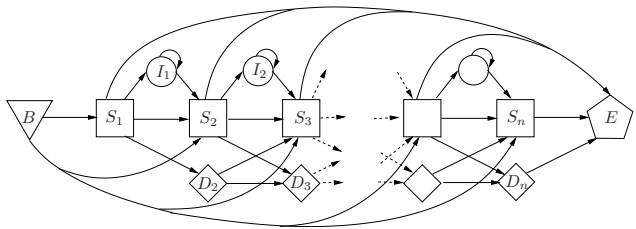}}
\caption{An example of a protein profile HMM architecture used by HMMER. The model contains $n$ positions plus a begin state ($B$) and end state ($E$). Each position contains a substitution ($S$) and a deletion state ($D$), with a possible insertion state ($I$) between two $S$-nodes. Allowed transitions are shown by arrows. To simulate local alignments, transitions $B\To S_i$ and $S_i\to E$, for any $S_i$, are permitted.}\label{fig:hmm}
\end{center}
\end{figure}

Two publicly available sources of curated alignments are the Pfam \citep{FMSGHLM06} and SUPERFAMILY \citep{GKHC01,WMVCG07} databases. In both, each MSA is represented by an HMM, which may be used for similarity searches. An HMM is a finite-state automaton, characterized by state-to-state transition probabilities and emission probabilities that generate hypothetical protein sequences. See Fig.~\ref{fig:hmm} for an example and the Appendix for more details.

The HMMER package \citep{Eddy98,Eddy03} uses the Viterbi algorithm \citep{DEKM98}, which finds the highest-scoring sequence of states in the HMM that produces the database sequence. The probability that a particular amino acid is emitted in a HMMER substitution state may be identified with the probability that it occurs in a corresponding position in a PSI-BLAST PSSM. On the other hand, HMMER allows position- and composition-specific gap scores, which model the probability that an insertion or deletion occurs at a particular position in an alignment.

With their greater gap cost flexibility, HMMs may be expected to have better retrieval accuracy than PSSMs. We attempt to quantify the effect of HMMER's use of more general gap parameters by separately examining the influence of position- and composition-specific gap scores. We also compare the retrieval accuracy of the PSSMs constructed using PSI-BLAST's iterative procedure to that of the HMMs provided by the Pfam and SUPERFAMILY collections. Our results may suggest some directions for improvements to PSI-BLAST, and the magnitude of the improvements one might expect.

We collected protein profile HMMs from Pfam and SUPERFAMILY. We then modified the profiles from each source to simulate different retrieval strategies, and used them as queries for HMMER and PSI-BLAST to search a set of sequences from the SCOP (structural classification of proteins) database \citep{MBHC95,AHCBHCM07}, which forms our `gold standard'. We use the results of the searches to evaluate and compare the retrieval performance of the search methods considered.

SCOP is a database of protein domains, classified by structure, function and sequence. Protein domains are classified into a hierarchy of class, fold, superfamily and family. Domains sharing the same superfamily are assumed to be homologous. For our testing purposes, we use the ASTRAL~40 \citep{CHWLKLB04} subset of SCOP (release 1.71), consisting of domain sequences that were filtered so that no two sequences share more than $40\%$ pairwise identity. ASTRAL has been used as the testing set in a number of performance evaluations of protein sequence comparison algorithms \citep{GB02,VGA04,PCGB05,YGASA06}.  

It is generally useful to evalute not only the difference in performance of two search methods, but also whether such a difference is statistically significant. A number of procedures have been proposed, mostly based on bootstrap resampling with replacement \citep{GB02,PCGB05}. In this context, \citet{GB02} observed that large superfamilies have an undue influence on the results, as the number of possible relationships grows quadratically with the number of members in a superfamily. They therefore proposed two weighting schemes that reduce the influence of large superfamilies. \citet{PCGB05} noted technical challenges in obtaining accurate variances for the weighted statistics and proposed an improved bootstrap. 

Our query sets, based on Pfam and SUPERFAMILY, contain several models for each SCOP-classified superfamily. Some superfamilies are overrepresented both in the query sets and in the ASTRAL database. We propose a different method than \citet{PCGB05} to address the difficulties associated with having superfamilies of different sizes. 
Our strategy is to sample without replacement $3/4$ of the superfamilies and then select a single model for each superfamily in any given query set. Hence, each sample contains no more than a single profile from each superfamily and therefore captures the most distant relationships among queries.

\section{Materials and Methods}

\subsection*{Software tools}

For HMM-based queries, we used the HMMER package (version 2.3.2) \citep{Eddy98,Eddy03}, which is also used internally by Pfam. Local alignment between a sequence and an HMM is allowed by the nonzero probabilities of entering match nodes directly from the begin state, as well as moving directly to the end state from them (Fig.~\ref{fig:hmm}). The statistical significance of each alignment score is estimated using an assumed extreme value distribution, with model-specific parameters. The final E-value, adjusted for model and sequence composition, is used to rank the hits. Another popular HMM platform is SAM \citep{HK96,BHK97,KKSH05}, which is used by SUPERFAMILY. We used HMMER rather than SAM for all our HMM-based queries because the programs' retrieval performances were shown to be comparable \citep{MG02,WS05} and because the SUPERFAMILY models were available in HMMER format.

For PSSM-based queries, we used PSI-BLAST (version 2.2.17) \citep{AMSZZML97}. The statistics of PSI-BLAST scores are based on the extreme value distribution \citep{Gumbel58} with a correction for finite sequence length. The statistical significance of each database hit is refined by taking into account its composition as well as that of the PSSM \citep{SAMSSWKA01}. 

PSI-BLAST allows one to start a search from a `checkpoint' file containing a PSSM saved from an earlier PSI-BLAST run, or built by other means. In addition to a PSSM, PSI-BLAST requires gap penalties as input: a gap opening cost and a gap extension cost. The choice of gap penalties is restricted to a few values because the parameters required to produce accurate statistics are precomputed using large-scale simulations. For both HMMER and PSI-BLAST runs, we used the standard search exectutables with their default settings.

\subsection*{Query sets}

Following \citet{WS05}, we constructed a query set of Pfam (release 22.0) models by identifying all Pfam-A models that were cross referenced by Pfam with an identifier in SCOP~1.71, and mapping the cross-referenced SCOP identifier to a SCOP superfamily. We did not consider models that have multiple domains mapping to different superfamilies.

We filtered the resulting set of Pfam models using two additional rules. First, any model mapping to a SCOP superfamily that had fewer than four members in ASTRAL~40 was removed from further consideration, to avoid superfamilies with a small number of members from disproportionally influencing the results. Next, we examined the MSA used to generate the Pfam profile and kept only those families whose MSA contained at least 10 sequences and had an average sequence length of at least 30 amino acids. Our final Pfam query set contained 703 Pfam models representing 299 superfamilies. We used the profiles from the \texttt{Pfam\_fs} set, built for local/local alignment.

Our second query set consisted of all 6729 models from the SUPERFAMILY database (release 1.69) that belonged to the 299 superfamilies in the Pfam query set. These models were also built for local/local alignment. The above query sets, paired with HMMER, formed our first two search methods, which we named \hmmerOF\ (\underline{H}MM, `\underline{o}riginal', P\underline{f}am) and \hmmerOU\ (\underline{H}MM, `\underline{o}riginal', S\underline{U}PERFAMILY).

\begin{table}[!h]
\begin{center}
\begin{tabular}{ll}\hline
Name & Description \\ \hline
\hmmerO & Original HMM dataset. \\
\hmmerM & HMMs, background insertion emission probabilities\\
\hmmerG & HMMs, constant state transitions and background insertion\\
&\qquad  emissions\\
\psiblastO & PSSMs, converted from original HMMs.\\
\psiblastC & PSSMs, from $5$ PSI-BLAST iterations over \texttt{nr} using profile\\
& \qquad consensus seeds\\
\psiblastS & PSSMs, from $5$ PSI-BLAST iterations over \texttt{nr} using SCOP\\ 
& \qquad domain sequence seeds\\ \hline
\end{tabular}
\caption{Nomenclature of query sets.
As shown in this table, the first two letters of the abbreviations of various search strategies denote the type of profile (HMM or PSSM), and the method of construction. The third letter is optionally appended to show the database of origin (\textsf{F} for Pfam, \textsf{U} for SUPERFAMILY). 
\label{tbl:nomen}}
\end{center}
\end{table}

The second pair of search methods, called \hmmerMF\ and \hmmerMU\ (see Table \ref{tbl:nomen} for an outline of all search methods), was constructed by taking the HMMs from \hmmerOF\ and \hmmerOU, respectively, and replacing all emission scores for each insert state with $0$. This is equivalent to setting all insertion emission probabilities to the background probabilities.

We constructed the third pair of search methods, called \hmmerGF\ and \hmmerGU\ by taking the HMMs from \hmmerMF\ and \hmmerMU, respectively, and adjusting the state transition probabilities to correspond to those implied by the affine gap penalties used by PSI-BLAST (see Appendix for a detailed explanation).

Let $\alpha$ denote the gap opening cost and $\beta$ the gap extension cost, in bits. We used the default penalty of PSI-BLAST, which is $11$ ($\alpha=5.040$ bits) for gap opening and $1$ for gap extension ($\beta=0.458$ bits). This scale was chosen to match the scale of BLOSUM62 \citep{HH92}, the default scoring matrix of BLAST.  

For each position $m$ of an HMM, we left the probabilities $P(B\To S_m)$ and $e_m = P(S_m\To E)$ unchanged and set the remaining transition probabilities as follows:
\begin{align}
P(D_m\To D_{m+1}) = P(I_m\To I_m) &= \nu,\\
P(D_m\To S_{m+1}) = P(I_m\To S_{m+1}) &= 1-\nu,\\
P(S_m\To D_{m+1}) = P(S_m\To I_m) &= {\mu (1-e_m) \over 1+ 2\mu - \nu},\\
 P(S_m\To S_{m+1}) &= {(1-e_m)(1-\nu )\over 1+ 2\mu - \nu},
\end{align}
where $\mu = 2^{\alpha+\beta}$ and $\nu=2^\beta$.
The probabilities were read from HMMER files, converted from scores, modified and written back as scores, as per HMMER convention \citep{Eddy03}. After modification, the HMMER statistical parameters of each HMM of \hmmerMF, \hmmerMU, \hmmerGF\ and \hmmerGU\ were recalibrated.

The remaining search methods used PSI-BLAST with default gap penalties. \psiblastOF\ and \psiblastOU\ used PSSMs derived from \hmmerOF\ and \hmmerOU, respectively, by taking the match state emission probabilities and writing them in PSI-BLAST checkpoint format. \psiblastCF\ and \psiblastCU\ used PSSMs obtained using the standard PSI-BLAST iterative procedure. We obtained the consensus (most likely) sequences of \psiblastOF\ and \psiblastOU\ profiles and used them as seeds for the initial searches, running $5$ iterations in total against \texttt{nr}, the database of non-redundant protein sequences maintained by NCBI (frozen on Apr 11, 2007) \citep{WBBBCC07}. 

The final search method, named \psiblastSU\ used the same construction procedure as 
\psiblastOU\ except that the SCOP sequences associated with SUPERFAMILY models were used as PSI-BLAST seeds instead of profile consensus sequences. 

\begin{figure}[!t]
\begin{center}
\begin{tabular}[t]{lr}
\multicolumn{1}{l}{\mbox{\bf (a)}} &
        \multicolumn{1}{l}{\mbox{\bf (c)}} \\ [-0.05cm]
\scalebox{0.45}[0.45]{\includegraphics{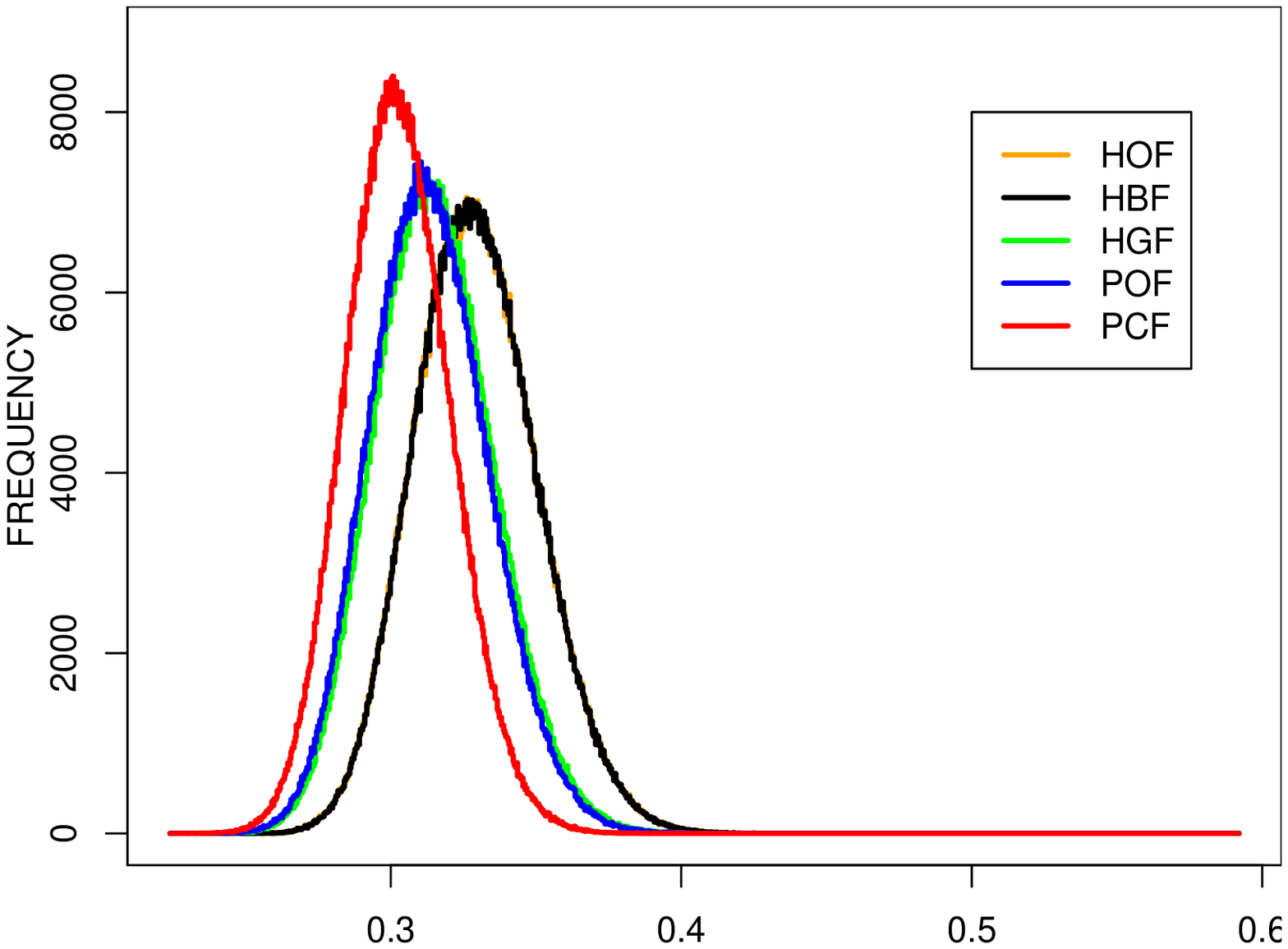}} &
\scalebox{0.45}[0.45]{\includegraphics{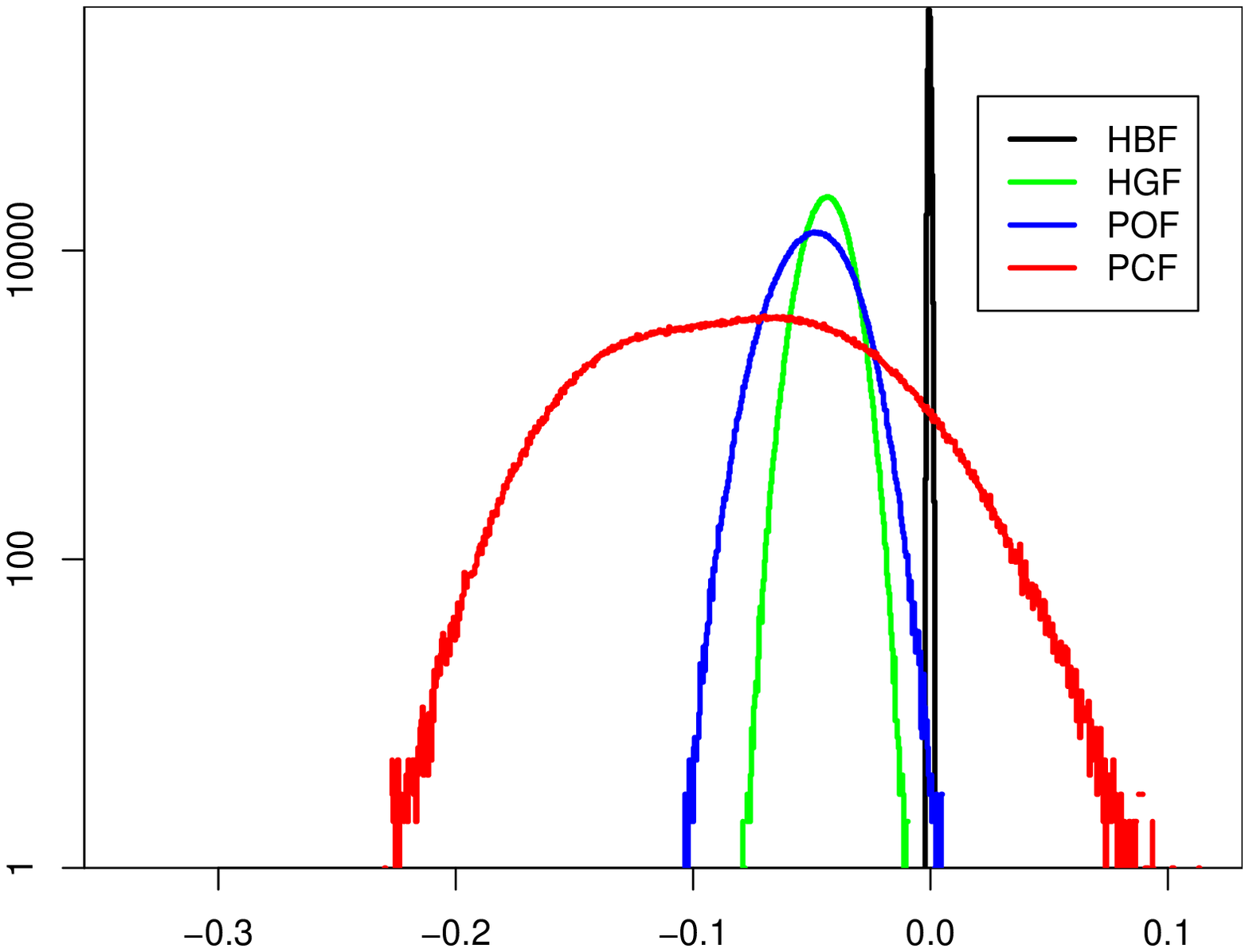}}\\ [-0.20cm]

\multicolumn{1}{l}{\mbox{\bf (b)}} &
        \multicolumn{1}{l}{\mbox{\bf (d)}} \\ [-0.1cm]
\scalebox{0.45}[0.45]{\includegraphics{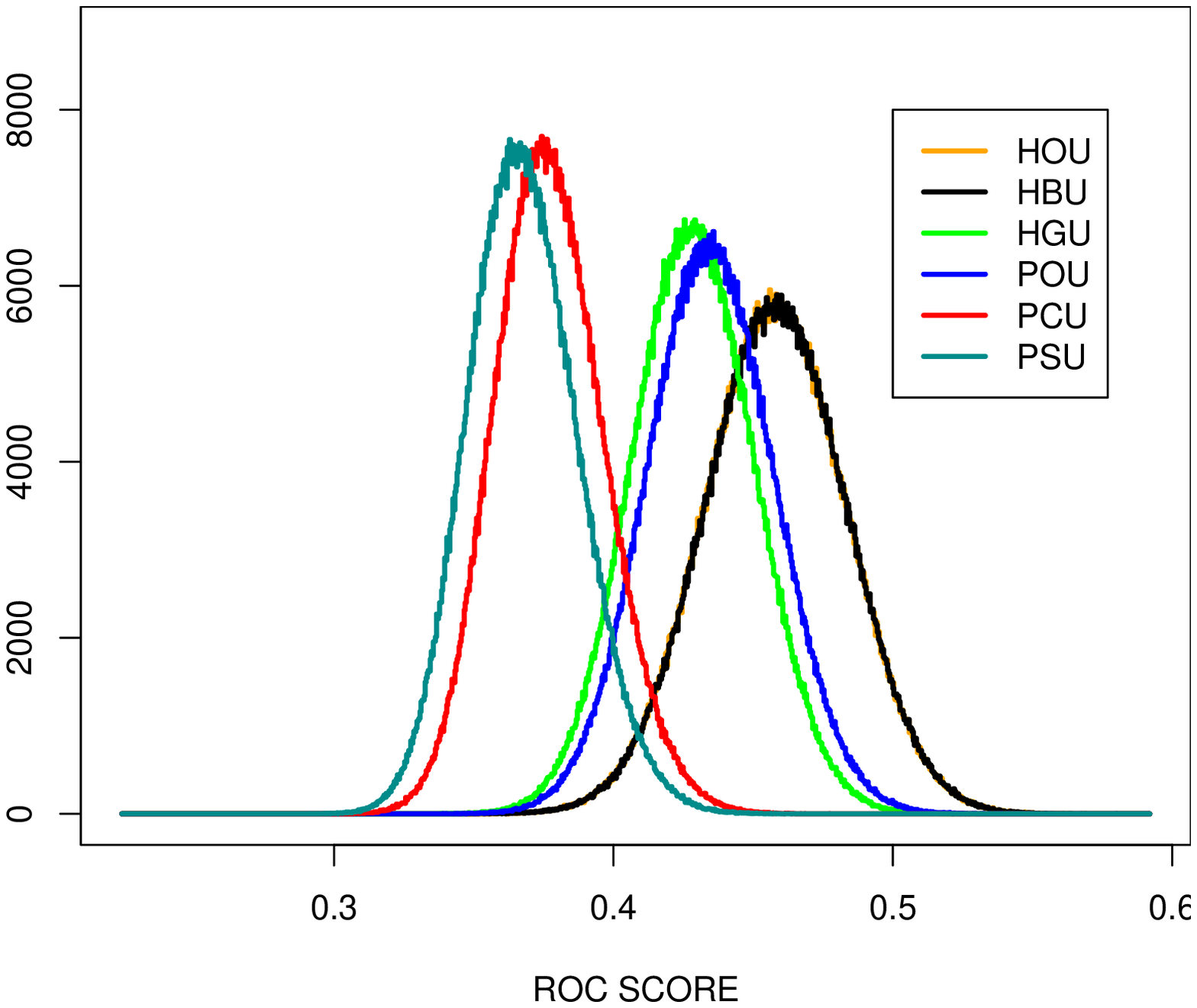}} &
\scalebox{0.45}[0.45]{\includegraphics{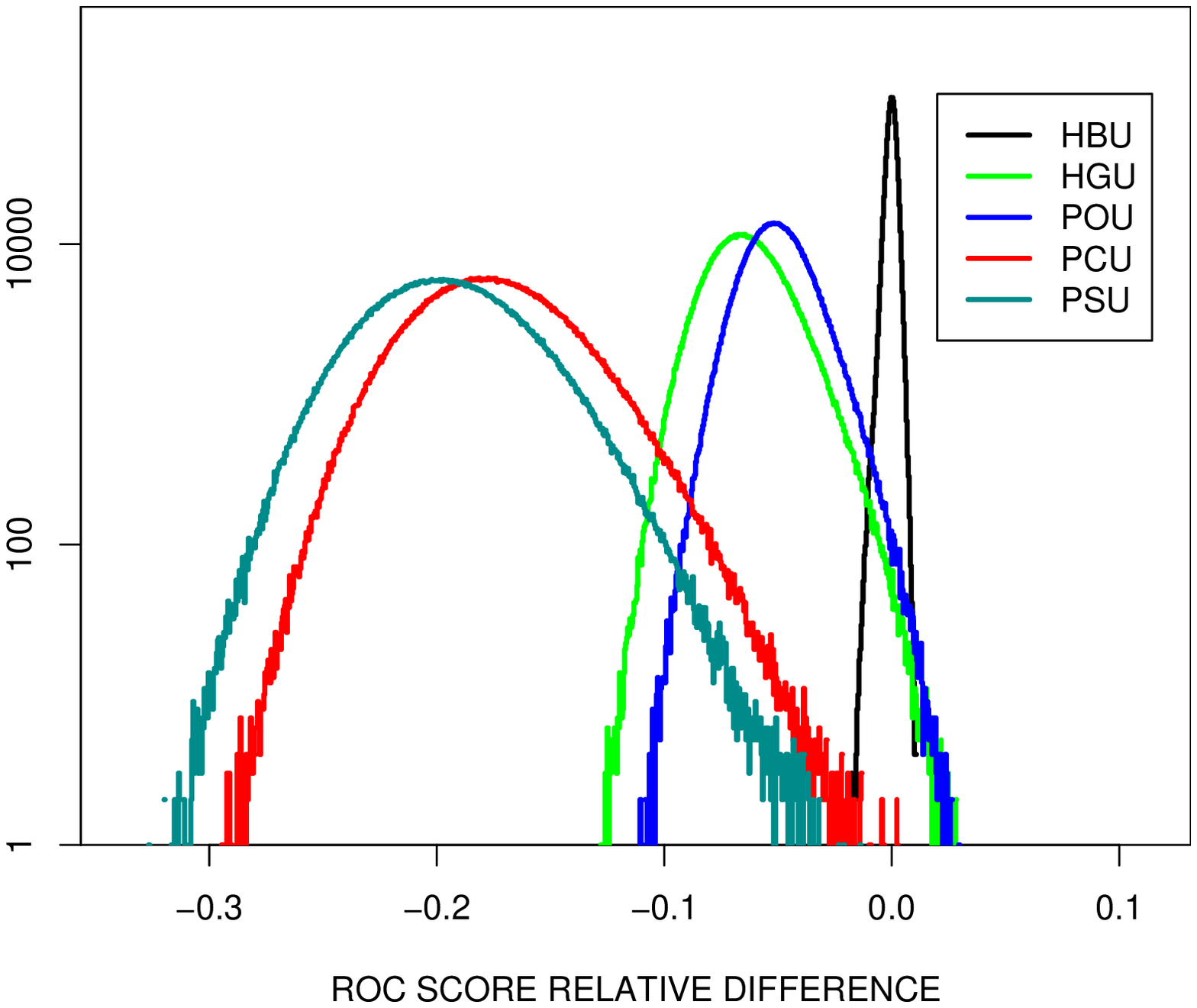}}\\
\end{tabular}
\caption{ROC score statistics of one million samples. In each sample, $224$ superfamilies are first randomly chosen from $299$ superfamilies. A representative query profile is then randomly selected from
 each chosen superfamily. ROC score histograms from using Pfam HMMs (a) and SUPERFAMILY HMMs (b) show appreciable difference in average ROC scores for each search method tested: SUPERFAMILY HMMs always perform better. Note that in panels (a) and (b), the curve for \hmmerO\ is completely covered by that for \hmmerM. Using \hmmerOF\ and \hmmerOU\ as baselines,  
the values of $\RRSD{224}$ (measurement at $1$ EPQ) between various methods and  the baselines are computed for each sample. The 
resulting histograms are shown in panels (c) and (d).
}\label{fig:mainres}
\end{center}
\end{figure}

\subsection*{Performance evaluation}

As described above, our query sets contained no profiles assigned to more than one SCOP superfamily. Each pair $\mathbf{p},\mathbf{s}$, where $\mathbf{p}$ is a query profile and $\mathbf{s}$ is an ASTRAL sequence, was classified as similar (`positive') if $\mathbf{s}$ belongs to the superfamily associated with $\mathbf{p}$,  and not similar (`negative') otherwise. For every query $\mathbf{p}_k$ from a set of queries, denote by $N_p(\mathbf{p}_k)$ the number of ASTRAL~40 sequences belonging to the same superfamily as $\mathbf{p}_k$ (i.e. the total number of positives for $\mathbf{p}_k$) and let $N_p=\sum_k N_p(\mathbf{p}_k)$.

Comparing each query profile to the ASTRAL~40 database, we retrieved a number of sequences ranked according to their E-values. These sequences were classified as true or false positives. For a given search strategy, after merging the results for the whole set of queries, we obtain the (step) functions $p(E)$ and $f(E)$ giving respectively the cumulative numbers of true and false positives with E-value $E$ or smaller. The function $p$ can also be expressed as a function of $f$, the number of false positives and the graph of $p(f)$ versus $f$ is called the ROC (receiver operating characteristic) curve \citep{HM82,GR96,HH02}. The same curve can be displayed as a coverage vs. error-per-query (EPQ) or CVE plot.

Our main performance statistic is the (truncated) ROC score. Given a number of false positives $F$, the $\ROC{F}$ score is defined by
\begin{equation}
\ROC{F} = \frac{1}{F\cdot N_p}\int_0^F p(f)\ df.
\end{equation}
It represents the accuracy of the search method (given a set of queries) for a given number of false positives. To compare two search methods $M_1$ and $M_2$ we compute their relative $\ROC{F}$ score difference, denoted $\RRSD{F}$, defined by
\begin{equation} 
\RRSD{F}(M_1,M_2) \equiv \frac{\ROC{F}(M_1) - \ROC{F}(M_2)}{\ROC{F}(M_2)}. \label{roc_reldiff}
\end{equation}

To overcome the aforementioned problems associated with overrepresentation of large superfamilies, we sampled according to the superfamily classification. For each sample we randomly picked 224 out of 299 superfamilies (leaving $1/4$ out) without replacement. Then, we selected one representative profile for each superfamily to form a sample query set. Search methods using the profiles originating from the same source (Pfam or SUPERFAMILY) used the same samples so that their performances could be compared for each sample. Our main statistic is the $\RRSD{224}$ per sample, which measures performance at $1$ EPQ or less. It allows a fair comparison of search methods.

\section{Results}

\begin{table}[!t]
{\tiny \begin{tabular}{|c|rrr|rrr|rrr|rrr|rrr|rrr|}
\hhline{----------------~~~} 
\textbf{(a)}& & \hmmerOF & & & \hmmerMF & & & \hmmerGF & & & \psiblastOF & & & \psiblastCF &  \\
\hhline{----------------~~~}
\hmmerOF & 0.0 & 0.0 & 0.0 & -0.1 & 0.0 & 0.1 & 2.9 & 4.5 & 6.3 & 2.4 & 5.1 & 8.1 & -0.6 & 8.5 & 19.5 \\ 
\hmmerMF & -0.1 & -0.0 & 0.1 & 0.0 & 0.0 & 0.0 & 2.9 & 4.5 & 6.3 & 2.4 & 5.1 & 8.1 & -0.6 & 8.5 & 19.4 \\ 
\hmmerGF & -5.9 & -4.3 & -2.8 & -5.9 & -4.3 & -2.8 & 0.0 & 0.0 & 0.0 & -1.7 & 0.5 & 3.0 & -4.9 & 3.8 & 14.1 \\ 
\psiblastOF & -7.5 & -4.8 & -2.3 & -7.5 & -4.8 & -2.4 & -3.0 & -0.5 & 1.8 & 0.0 & 0.0 & 0.0 & -5.1 & 3.2 & 13.1 \\ 
\psiblastCF & -16.3 & -7.8 & 0.6 & -16.3 & -7.8 & 0.6 & -12.3 & -3.7 & 5.1 & -11.6 & -3.1 & 5.3 & 0.0 & 0.0 & 0.0 \\ 
\hhline{----------------~~~}
\multicolumn{2}{c}{}\\[0.25cm]

\hline
\textbf{(b)} & & \hmmerOU & & & \hmmerMU & & & \hmmerGU & & & \psiblastOU & & & \psiblastCU &  & & \psiblastSU & \\ 
\hline
\hmmerOU & 0.0 & 0.0 & 0.0 & -0.4 & -0.0 & 0.4 & 2.9 & 6.9 & 10.2 & 2.0 & 5.3 & 8.2 & 12.4 & 21.4 & 30.3 & 15.0 & 24.5 & 34.2 \\ 
\hmmerMU & -0.4 & 0.0 & 0.4 & 0.0 & 0.0 & 0.0 & 2.9 & 6.9 & 10.2 & 2.1 & 5.3 & 8.2 & 12.4 & 21.4 & 30.3 & 15.0 & 24.5 & 34.2 \\ 
\hmmerGU & -9.3 & -6.4 & -2.8 & -9.3 & -6.4 & -2.8 & 0.0 & 0.0 & 0.0 & -4.1 & -1.4 & 1.4 & 7.0 & 13.6 & 20.8 & 9.6 & 16.4 & 24.3 \\ 
\psiblastOU & -7.6 & -5.0 & -2.0 & -7.6 & -5.1 & -2.0 & -1.4 & 1.5 & 4.3 & 0.0 & 0.0 & 0.0 & 8.1 & 15.3 & 22.8 & 10.7 & 18.2 & 26.3 \\ 
\psiblastCU & -23.3 & -17.6 & -11.0 & -23.3 & -17.7 & -11.0 & -17.2 & -12.0 & -6.6 & -18.6 & -13.3 & -7.5 & 0.0 & 0.0 & 0.0 & -1.5 & 2.5 & 7.2 \\ 
\psiblastSU & -25.5 & -19.6 & -13.0 & -25.5 & -19.7 & -13.1 & -19.5 & -14.1 & -8.8 & -20.8 & -15.4 & -9.7 & -6.7 & -2.4 & 1.5 & 0.0 & 0.0 & 0.0 \\ \hline 
\end{tabular}}
\caption{Summary of statistics of $\RRSD{224}$ between every pair search strategies using the same source.
In Fig. \ref{fig:mainres}~(c) and (d), \hmmerOF\ and \hmmerOU\ were used as the baselines for Pfam and SUPERFAMILY search strategies, respectively, and the histograms of $\RRSD{224}$ relative to the baselines are shown. It is impractical to show such histograms for all possible baselines. However, for each pair of search strategies, we may sort (in ascending order) their one million values of $\RRSD{224}$ and record the corresponding $\RRSD{224}$ value at various designated percentiles. In the table, there are three numbers in a row for any given pair of search strategies. As an example, the numbers  $2.9$,  $4.5$ and $6.3$, associated with $M_1=\hmmerMF$ and $M_2=\hmmerGF$, are located in the row labelled by $\hmmerMF$ and within the column headed by $\hmmerGF$. Those numbers, when divided by $100$, have the following interpretation: the leftmost corresponds to the $\RRSD{224}$ value at the $2.5$-th percentile, the middle to the median and the rightmost to the $97.5$-th percentile. Subtable (a) records the numbers associated with Pfam search methods, while subtable (b) documents those associated with the SUPERFAMILY strategies tested.
\label{tbl:reldiff}} 
\end{table}

Fig.~\ref{fig:mainres} shows the distributions of $\ROC{224}$ scores and their relative differences ($\RRSD{224}$) \textit{per sample} with respect to \hmmerO\ for all query sets. Comparison of Fig.~\ref{fig:mainres} (a) and (b) shows that, in general, the strategies using profiles from SUPERFAMILY perform better than those using Pfam profiles. In terms of relative difference (Fig.~\ref{fig:mainres}\ (c,d), Table \ref{tbl:reldiff}), using both Pfam and SUPERFAMILY profiles, original HMMs (\hmmerO) perform significantly better than all other query sets except \hmmerM. There is no perceivable difference between \hmmerM\ and \hmmerO. There is also no significant difference between \hmmerG\ and \psiblastO.

In the case of PSSMs, \psiblastOU\ gives better performance than \psiblastCU\ and \psiblastSU, but there is no significant difference between \psiblastOF\ and \psiblastCF, although \psiblastCF\ shows a large variance in performance. In a number of cases, a \psiblastCF\ sample even outperforms the corresponding \hmmerOF\ sample. The relative ROC score difference between \psiblastCU\ and \psiblastSU\ is slightly positive, but not significantly so.

\begin{figure}[h!]
\begin{center}
\scalebox{0.5}[0.5]{\includegraphics{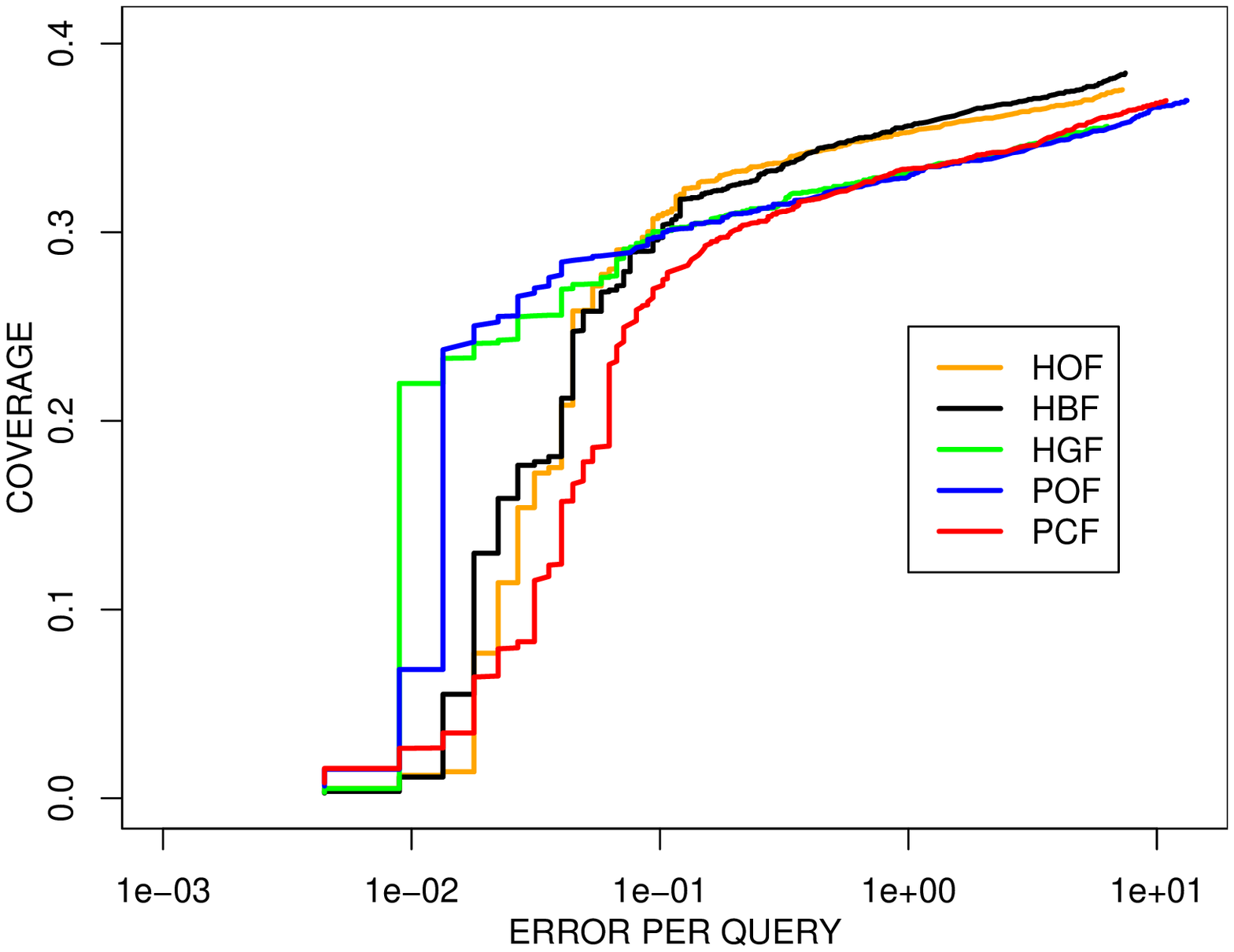}}
\vskip 0.05cm
\scalebox{0.5}[0.5]{\includegraphics{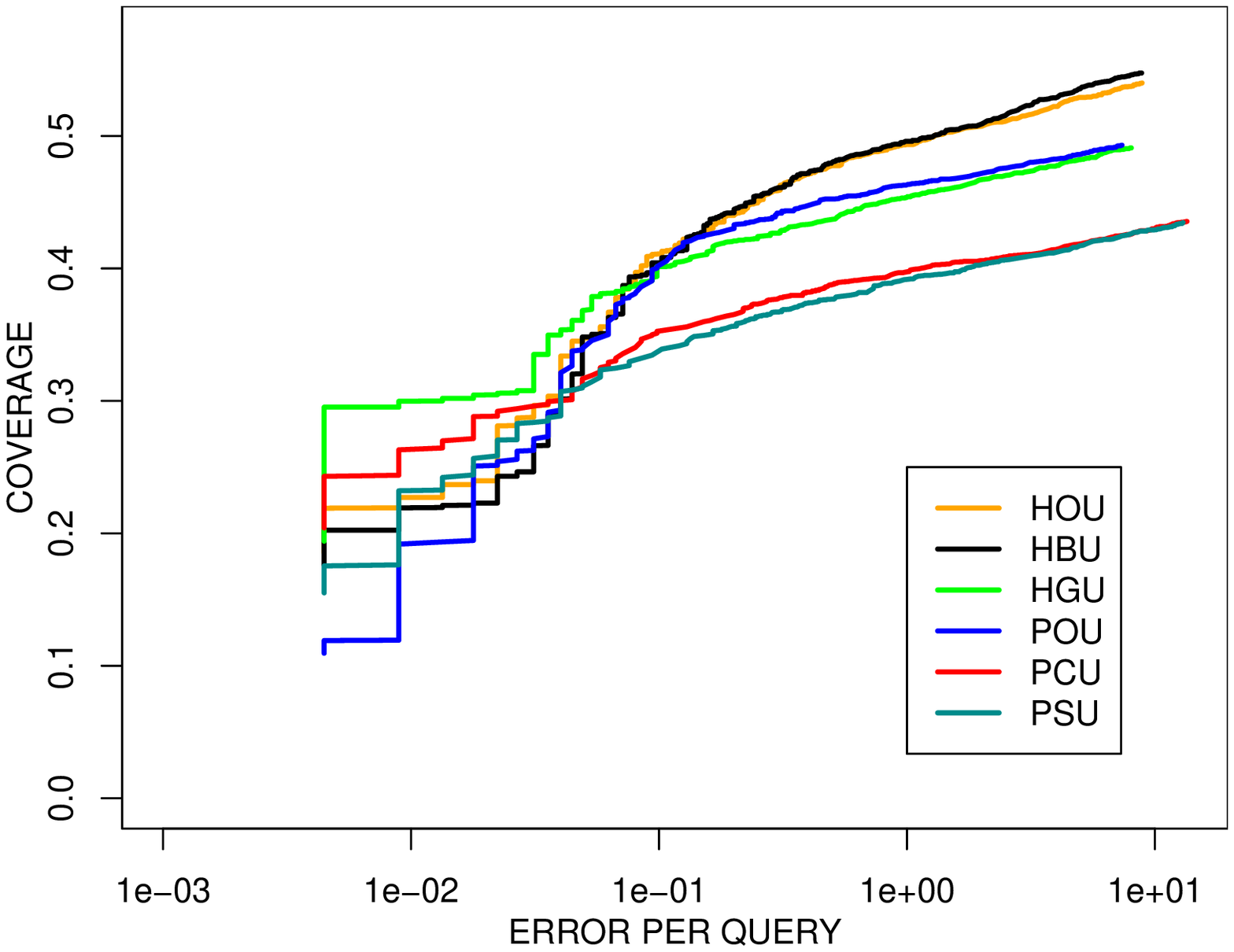}}
\caption{Example CVE curves for various search strategies based on Pfam (top) and SUPERFAMILY (bottom) profiles. Each curve shown is a representative that corresponds to a sample with $\ROC{224}$ score equal to the median of 1,000,000 samples.}\label{fig:roccurves}
\end{center}
\end{figure}

Using profiles from Pfam (SUPERFAMILY), we observed two (three) clusters of search strategies that performed equivalently based on $\RRSD{224}$ (Fig.~\ref{fig:mainres}\ (c,d)). This trend in performance is supported by Fig.~\ref{fig:roccurves}, which displays examples of CVE curves for all alignment methods tested. The samples associated with these CVE curves have the median $\ROC{224}$ score. 

\section{Discussion and Conclusion}

The clear separation in retrieval performance between the SUPERFAMILY and Pfam profiles could be explained by the fact that the former are based on ASTRAL sequences, which form our testing set as well. In contrast, Pfam models are based on a variety of sequence sources and were not trained on ASTRAL. Hence, a degree of overfitting the SUPERFAMILY models to the testing set, as well as the fact that ASTRAL is structure based, may explain the overall differences in performance. 

Another interesting observation is that CVE curves (Fig.~\ref{fig:roccurves}) cross at low EPQ and form distinct clusters above 0.5 EPQ. Due to small sample size, the coverage at low EPQ is expected to have a larger uncertainty, thus the crossing of CVE curves there is anticipated. At moderate EPQ, the distinct clusters indicate that the relative retrieval efficiency is not influenced by the choice of EPQ.

On both testing collections, we have observed almost no difference in performance between the original HMMs (\hmmerO) and the models derived from them having insertion emission probabilities reset to the background (\hmmerM). Examining the models in HMMER format, we found that the insertion emission distributions were almost constant over all the positions, with the common distribution being slightly biased in favor of hydrophilic amino acids. The average relative entropy between this distribution and the background distribution is very small (0.037 bits for Pfam, 0.005 bits for SUPERFAMILY), explaining the very small effect of the insertion emissions on the retrieval performance. Note that SUPERFAMILY models had higher overall probabilities of entering a gap state and hence showed a larger influence of insertion emissions than Pfam models (Fig.~\ref{fig:mainres} (c,d)).

In addition, an insertion emission distribution biased in favor of hydrophilic amino acids may not be appropriate for all positions within proteins: it implicitly assumes the globular protein structure, with hydrophobic core and hydrophilic surface. Finally, from an information theoretic point of view, it is very difficult to reliably estimate insertion emission probabilities. In particular, if one wishes to establish an emission model whose  emission probabilities are similar to those of the background and wants to confidently distinguish those two sets of probabilities, it is necessary to have  a large amount of data. The following example illustrates this point.

In the Pfam insertion emission model, Leucine's emission probability, $0.0676$, has the largest deviation compared to the background $0.0934$. Consider a simple coin tossing experiment where the probability of seeing a leucine (head) is $p=0.0676$ and the probability of seeing any other amino acid (tail) is $1-0.0676$. One may ask how many tosses (number of amino acids present in a gap column of an MSA) are needed in order to confidently rule out the possibility that the probability is $0.0934$.  It is well known that a binomial distribution  in the large number limit becomes a Gaussian. In our example, the probability of observing $k$ heads out of $n$ tosses becomes
\[
C^n_k p^k (1-p)^{n-k} \Delta k{\approx}
\sqrt{n/p(1-p)\over 2 \pi}\,\, e^{-n({k\over n} -p)^2/2p(1-p) } d({k \over n})\; .
\]
To reject with $85\%$ confidence the value of $0.0934$ as the probability of seeing a head, the absolute difference between the two probabilities, $0.0934$ and $0.0676$, must be greater than or equal to $1.02$ times the standard deviation, $\sqrt{p(1-p)/n}$. This leads to 
\[
0.0258 > 1.02 \sqrt{0.0847 \over n} \Rightarrow 
 n > 132 \; . 
\]
When applied to estimating insertion emission probabilities, this example implies one needs to have about $132$ amino acids in a gap column of a multiple alignment. This number seems large for columns associated with an insert state, as these columns normally have more gaps than amino acids.  On the other hand, we can confidently determine emission probabilities for columns that contain mostly amino acids and are therefore usually assigned to substitution states. Furthermore, the dominant amino acid in a match column often has very different observed and background frequencies. For example, consider a match column with $20\%$ leucine. The same calculation as above tells us that we need only eight or more amino acids in the match column to indicate a preferrence for leucine. Of course, considering the sub-dominant amino acids requires more entries in the match column.

Comparing \hmmerO\ to \hmmerG\ and \psiblastO, we see that profiles with position-dependent gap parameters have 5\% better retrieval performance (as measured by the median $\RRSD{224}$ value) than those with position-independent ones. This is an area where HMMs are clearly superior to the PSSMs with constant gap penalties, as used by PSI-BLAST. Hence, a possible direction for improvement of PSI-BLAST is to introduce position-dependent gap parameters. When interpreting this difference, one should note that we did not optimize the PSI-BLAST gap penalties, but use only the defaults. It is therefore possible that the performance of PSI-BLAST with a better set of gap opening and extension penalties would more closely match the performance of HMMs. Another possibility is to estimate and optimize gap parameters for each PSSM separately, at the time of its creation (that is, each PSSM would still carry a single, position independent, gap opening and gap extension penalty, but they would not be input beforehand but estimated from the data). The practical problem with these suggested improvements is that the statistical parameters for position-specific gap penalties cannot be quickly computed as yet, and one is therefore restricted to the costs for which the parameters have been precomputed. Another possibility is to modify PSI-BLAST to use the hybrid alignment algorithm \citep{YH01,YBH02}, which is probabilistic, naturally accepts PSSMs with position-specific gap costs, and has well-characterized, universal statistics.

It is not surprising that the performances of \hmmerG\ and \psiblastO\ show no significant difference because \hmmerG\ was designed to simulate the PSI-BLAST gap parameters in the HMM framework. Some differences still exist due to a fundamental difference between the underlying algorithms. First, although the score statistics for HMMER and PSI-BLAST are both based on the extreme value distribution, there are still differences in details. Second, PSI-BLAST alignments may have longer segments of ungapped alignment because the score associated with ungapped alignment is not reduced by the probability of entering another node. Some difference can also be explained by slightly different background probabilities in each case. Finally, local alignment is achieved through different mechanisms: PSI-BLAST alignments terminate when their accumulated score is maximal, while HMMER alignments terminate only when they hit the end state. Thus, HMMER alignments may tend to be more global with respect to the profile.

The difference in performance of PSI-BLAST using PSSMs constructed in different ways shows that focusing on profile construction as well as on position-specific gaps may yield significant improvement. In particular, the performance of PSSMs converted from HMMs (\psiblastO) versus those iteratively constructed (\psiblastC\ and \psiblastS) shows that a more carefully constructed profile may yield better performance, with the difference being more pronounced in SUPERFAMILY than in Pfam. The fact that the PSSMs obtained iteratively from \texttt{nr} based on SUPERFAMILY consensus seeds generally perform better than those originating from Pfam consensus seeds shows the importance of the choice of the initial seed sequence. This is further emphasised by the slightly better performance of the PSSMs based on the consensus sequence as seed (\psiblastCU) than the performance of those based on the seeds taken from ASTRAL (\psiblastSU). Hence, another possible way of improving PSI-BLAST would be to run one iteration using the normal scoring matrix and construct a profile as before, but then to rerun the search using the consensus sequence as the seed instead of proceeding into the iterative stage with the profile. In that way, a more `central' seed can be obtained, which, while not corresponding exactly to any sequence present in the dataset, may yield a more accurate profile for the iterative steps. Naturally, the choice of the weighting scheme for the multiple alignment used to obtain the consensus sequence or profile as well as the associated pseudocounts will also exert a significant influence on the result.

Finally, our methodology must be understood in the context of the small size of the testing suite. This does not present a significant problem when testing different parameter sets of the same alignment algorithm but when comparing different algorithms, it is essential to eliminate bias due to superfamily size. Our approach, based on sampling $3/4$ of the superfamilies without replacement, was designed with this aim in mind.

%
%

\section*{Funding}
This work was supported by the Intramural Research Program of the National Library of Medicine at National Institutes of Health. 

\section*{Acknowledgement}
We thank M. Wistrand and E. Sonnhammer for useful correspondence.


\section*{Appendix}
The connection between the transition probabilities of HMMs for sequence evolution
and the scoring function (scoring matrices and gap parameters) used in sequence comparison is elaborated in this appendix. Since such a connection has been sketched explicitly in earlier publications on hybrid alignment \citep{YH01,YBH02}, interested readers are encouraged to look into the original literature. We present a self-contained exposition here to save the reader some effort in reading through earlier papers, and to present a minor extension needed for aligning a protein sequence to a {\it local} HMM with explicit termination probabilities at its nodes. Note that keeping a nonzero termination probability  is how HMMER achieves local alignments. Hybrid alignments achieve a local alignment by taking the maximum of the log-odd ratios at each possible termination point, and hence do not need to deal explicitly with the termination probabilities of the HMMs.  

The fundamental idea of protein sequence comparison is rooted in the amino acid score (substitution) matrix,
 where the $(i,j)$th  entry 
\be
s_{ij} = {1\over \lambda} \ln \left[ {Q_{ij} \over p_i p_j} \right] 
\ee
is the log-odd ratio of the joint probability $Q_{ij}$ of amino acids $i$ and $j$ in 
 the target ensemble to the product of the background probabilities, $p_i$ and $p_j$, of the two amino acids. Here $\lambda$ is just a scale and is set to unity from this point on.  For a valid scoring matrix \citep{YWA03}, one has $p_i = \sum_j Q_{ij}$ and one may express $Q_{ij}$ as  $Q_{ij} = p_i T(j|i) = p_j T(i|j)$, with $T(j|i)$ being the probability for amino acid $i$ to mutate into amino acid $j$. In this case, we may also write 
\be
s_{ij} = \ln \left[ {T(j|i) \over p_j} \right], 
\ee
which may now be viewed as the log-odds ratio of a conditional emission probability to the background probability.

Extending this concept \citep{YH01,YBH02}, one may score the global relatedness (alignment) between two protein sequences, $\ba$ and $\bb$, the same way: using the log-odds ratio of $Q[\ba,\bb]$ to $P[\ba] P[\bb]$ (the background probability of generating a pair of random sequences $\ba$ and $\bb$). In terms of global relatedness, $Q[\ba, \bb]$ may be regarded again as $P[\ba] T[\bb | \ba ]$ and
\be
{ Q[\ba, \bb] \over P[\ba] P[\bb] } = { T[\bb | \ba] \over P[\bb ] }.
\ee
 Here $T[\bb |\ba]$ is the probability for sequence $\ba$ to mutate into sequence $\bb$. It is not hard to convince oneself that there are many different "ways" or "paths" for sequence $\ba$ to mutate into sequence $\bb$. In fact, it has been argued that the  usual {\it optimal} alignment corresponds to the {\it most probable} evolutionary path. In this context, the gap cost is related to the transition probabilities in and out of the insertion/deletion states of the HMM. 

A protein HMM consists of a number of nodes. Except at the begin node, each node $j$ allows two possible states, substitution (S) and deletion (D). The substitution state associated with node $j$ is characterized by the transition probability from $a_j$ to other amino acids. The deletion state is further divided into cases depending on its preceding state. In between two nodes, one can have an insertion (I) state. The transition probabilities from a given state to all other allowed states have to sum to one. Four transitions -- $S \To S$, $S\To D$, $D\To S$, and $D \To D$ --  will each advance the node index by one. Transition $S \To I$ and $I \To S$  combined together increase effectively the node index by $1$, while the transitions $I \To I$ and $D \To I$ (if allowed) do not change the node index at all. In many HMMs, such as the ones used by HMMER, the transitions between $I$ and $D$ states are strictly forbidden and we follow this rule in here to simplify our exposition. 

\begin{figure}[h!]
\begin{center}
\scalebox{0.70}{\includegraphics{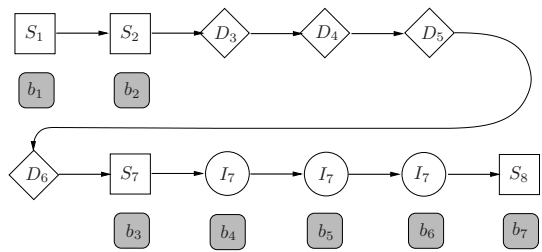}}
\caption{An example of a partial alignment between a profile HMM and a protein sequence. Note that in the text, the state preceding $S_1$ is assumed to be a substitution state.}\label{fig:align}
\end{center}
\end{figure}

Constrained by the probability conservation condition, the transition probabilities are usually made node-specific (or equivalently termed {\it position-specific}). Focusing on the substitution scores of protein HMMs, position-specific scoring simply means that the substitution states at different nodes may emit amino acids with different sets of probabilities.
 
As a concrete example, let us consider an alignment of a partial HMM model
 of eight nodes aligned with a sequence $\bb = [b_1, b_2, \ldots, b_7]$ of seven amino
 acids. Their alignment is shown in Fig.~\ref{fig:align}. The alignment score ${\sf S}$ is given by
\bea 
 {\sf S} &=& s_{1}(b_1)+ s_2(b_2) + g^D_{(3-6)} + s_7(b_3) \nonumber \\
 && + g^I_{7^+}([b_4,b_5,b_6])+ s_8(b_7) \label{score}
\eea
 where $s_i(b_j)$ represents the substitution score for amino acid $j$ at node $i$, $g^D_{(3-6)}$ represents the gap score associated with deleting nodes $3$ through $6$,  and $g^I_{7^+}([b_4,b_5,b_6])$ represents the gap score associated with inserting amino acids $b_4, b_5$ and $b_6$ between nodes $7$ and $8$ of the HMM. The superscript "+" associated with the $I$ state will be suppressed from this point on. The probability of occurrence associated with this alignment ${\mathcal A}$ may be written as 
\bea
T_{\mathcal A}[\bb | \ba]\!\!\!\! &=& \!\!\!\! P(S_0\To S_1) T_{a_1}(b_1) P(S_1 \To S_2) T_{a_2}(b_2) P(S_2\To D_3) \nonumber \\ 
 && \!\!\!\! P(D_3\To D_4) P(D_4\To D_5) P(D_5 \To D_6) P(D_6\To S_7) \nonumber  \\ 
 && \!\!\!\! T_{a_7}(b_3)  P(S_7 \To I_7) \tilde p(b_4) P(I_7\To I_7) \tilde p(b_5)\nonumber  \\ 
 && \!\!\!\! P(I_7\To I_7) \tilde p(b_6) P(I_7\To S_8) T_{a_8}(b_7), \nonumber
\eea  
 where $\tilde p(b)$ is the insertion probability of amino acid $b$ between nodes
 $7$ and $8$. Assuming that $P[\bb] = \prod_i p(b_i)$,  one obtains the ratio
\bea 
{T_{\mathcal A}[\bb | \ba] \over P[\bb]}\!\!\!\! &=& \!\!\!\!
P(S_0 \To S_1) { T_{a_1}(b_1) \over p(b_1)} P(S_1\To S_2) {T_{a_2}(b_2) \over p(b_2)}
 P(S_2\To D_3) \nonumber \\ 
 && \!\!\!\! P(D_3\To D_4) P(D_4\To D_5) P(D_5 \To D_6) P(D_6\To S_7) \nonumber  \\
&& \!\!\!\! {T_{a_7}(b_3) \over p(b_3)}P(S_7 \To I_7) {\tilde p(b_4) \over p(b_4)} P(I_7\To I_7) 
{\tilde p(b_5) \over p(b_5)}\nonumber  \\
&&\!\!\!\!  P(I_7\To I_7) {\tilde p(b_6) \over p(b_6)} P(I_7\To S_8) {T_{a_8}(b_7) \over p(b_7)}. \label{prob_ratio} 
\eea

Comparing (\ref{score}) and (\ref{prob_ratio}) and events of similar type yields the following mappings:
\bea
\exp [s_i(b_j)] &=& P(S_{i-1} \To S_i) {T_{a_i}(b_j) \over p(b_j)} \nonumber \\
\exp [g^D_{(3-6)}]&=&  P(D_3\To D_4) P(D_4\To D_5) P(D_5 \To D_6)\nonumber \\
  && \hspace{10pt} \times {P(S_2\To D_3) P(D_6\To S_7)\over P(S_6 \To S_7)} \nonumber \\ 
\exp [g^I_7([b_4,b_5,b_6]) &=& {P(S_7 \To I_7) P(I_7 \To S_8) \over P(S_7 \To S_8)}
 \left[ P(I_7 \To I_7) \right]^2  \nonumber \\
 && \hspace{10pt} \times {\tilde p(b_4) \over p(b_4)}
 {\tilde p(b_5) \over p(b_5)} {\tilde p(b_6) \over p(b_6)} \; . \label{score_prob_relation}
\eea

Frequently HMMs take $P(D_{i-1} \To D_i)$ and $P(I_j \To I_j)$ each to be a constant. In this case, the ratio ${P(S_7 \To I_7) P(I_7 \To S_8) \over P(S_7 \To S_8)P(I\To I)}$ contributes to a  position specific gap opening cost and the ratio ${\tilde p(b) \over p(b)}$ contributes to  a composition-dependent insertion cost. The quantity $P(I\To I)$ contributes to the insertion gap extension cost.  If one keeps emission probabilities  $T_{a_i}(b)$ node-dependent, but demands that all the state-to-state transition  probabilities be node-independent, one essentially has a PSSM with uniform affine gap costs, although possibly with composition-specific insertion gap costs if $\tilde p$ is chosen to be different from the background $p$. 

Since the transition probabilities are constrained by the respective conservation conditions, and those probabilities are related to the scoring function through (\ref{score_prob_relation}), the substitution and gap scores are no longer independent if one wishes to have a  probabilistic interpretation. We now turn to the relationship among score parameters when the state to state transition probabilities are node-independent constants.  
 Let $\eta \equiv P(S \To S)$, $\mu^{D1} \equiv P(D\To S)/P(S\To S)$, $\mu^{D2} \equiv P(S\To D)$,
$\mu^{I1} \equiv P(I \To S)/P(S\To S)$, $\mu^{I2} \equiv P(S \To I)$, $\nu^I \equiv P(I\To I)$,
 and $\nu^D \equiv P(D \To D)$. Because $\mu^{I1}$($\mu^{D1}$) and $\mu^{I2}$($\mu^{D2}$) always appear together as a product, we further define $\mu^I \equiv \mu^{I1}\mu^{I2}$ ($\mu^D \equiv \mu^{D1} \mu^{D2}$). 
The probability conservation condition then 
 demands that 
\bea
\eta + \mu^{I2} + \mu^{D2} &=& 1 \label{Sout}\\
\eta \mu^{I1} + \nu^I &=& 1 \label{Iout}\\
\eta \mu^{D1} + \nu^D &=& 1. \label{Dout}
\eea
 Treating $\nu^I$, $\nu^D$, $\mu^I$ and $\mu^D$ as fixed parameters allows us 
 to express $\eta$, $\mu^{I2}$, and $\mu^{D2}$ in terms of $\nu$s and $\mu^{I(D)}$. 
 To do so, we multiply (\ref{Iout}) by $\mu^{I2}$ and multiply (\ref{Dout}) by $\mu^{D2}$.
 Together with (\ref{Sout}), we have three linear equations with three unknowns: $\eta$, $\mu^{I2}$,
 and $\mu^{D2}$. Solving these equations yields
\bea
\eta &=& {1\over 1+ \mu^I /(1-\nu^I) + \mu^D / (1-\nu^D)} \nonumber \\
\mu^{I2} &=& {\mu^I/(1-\nu^I) \over 1+ \mu^I /(1-\nu^I) + \mu^D / (1-\nu^D)} \nonumber\\
\mu^{D2} &=& {\mu^D/(1-\nu^D) \over 1+ \mu^I /(1-\nu^I) + \mu^D / (1-\nu^D)}. \nonumber 
\eea
For the case $\mu^D = \mu^I = \mu$ and $\nu^D = \nu^I = \nu$, these expressions simplify to 
\bea
\eta &=& {1-\nu \over 1+ 2\mu - \nu} \nonumber \\
\mu^{D2} = \mu^{I2} &=& {\mu \over 1+ 2\mu - \nu}. \nonumber
\eea
Note that with this notation, we may rewrite (\ref{score_prob_relation}) as
\bea
\exp [s_i(b_j)] &=& \eta {T_{a_i}(b_j) \over p(b_j)} \nonumber \\
\exp [g^D_{(3-6)}]&=& \mu \nu^3 \nonumber \\ 
\exp [g^I_7([b_4,b_5,b_6]) &=& \mu \nu^2 \times {\tilde p(b_4) \over p(b_4)}
 {\tilde p(b_5) \over p(b_5)} {\tilde p(b_6) \over p(b_6)}. \nonumber
\eea
It becomes evident that $\ln(\mu/\nu)$ corresponds to the gap opening score
while $\ln(\nu)$ corresponds to the gap extension score, and $\ln(\tilde p(b)/p(b))$
becomes an additional composition-specific insertion score.

In HMMER, the local alignment is terminated by going into the end state, and the end state can be reached only from substitution states. In this context, the probability conservation  equations (\ref{Iout}) and (\ref{Dout}) remain unchanged. However, we may allow a node-specific termination probability from the $S$ state. This requires the introduction of a position index for the other transition probabilities. Let $\eta_m \equiv P(S_m \To S_{m+1})$,
 $e_m \equiv P(S_m \To E)$, $\mu_m^{D1} \equiv P(D_m\To S_{m+1})/P(S_m\To S_{m+1})$, 
 $\mu_m^{D2} \equiv P(S_m\To D_{m+1})$,
$\mu_m^{I1} \equiv P(I_m \To S_{m+1})/P(S_m\To S_{m+1})$, $\mu_m^{I2} \equiv P(S_m \To I_m)$.
 However, note that $P(I_m \To S_{m+1})$ should remain the same, because there is no direct transition $ I_m \To E$. Thus, we may still keep both $\mu_m^{D1} \mu_m^{D2} = \mu_m^{I1}\mu_m^{I2} = \mu$
  and $\nu_m^D = \nu_m^I = \nu$ as constants. The probability conservation condition then yields
\bea
\eta_m + \mu_m^{I2} + \mu_m^{D2} + e_m &=& 1 \label{Sout.m}\\
\eta_m \mu_m^{I1} + \nu &=& 1 \label{Iout.m}\\
\eta_m \mu_m^{D1} + \nu &=& 1, \label{Dout.m}
\eea
the solution of which is  
\bea
\eta_m &=& {(1-e_m)(1-\nu )\over 1+ 2\mu - \nu} \nonumber \\
\mu_m^{D2} = \mu_m^{I2} &=& {\mu (1-e_m) \over 1+ 2\mu - \nu}. \nonumber
\eea
Although $\mu_m^{D2}$ and $\mu_m^{I2}$ are decreased,
$\mu_m^{D2}\mu_m^{D1} $ and $\mu_m^{I2}\mu_m^{I1}$ are kept the
same as before. As a consequence, the only change is that the
substitution score at each node is reduced by a node-specific constant
$\ln [1/(1-e_m)]$ when it is not preceded by a gap state. If an
alignment has deletion at node $m$ followed by $k$ more substitutions
from node $m+1$ to node $m+k$, then the substitution score reduction
starts only at node $m+2$ and persists to node $m+k$.

\end{document}